\documentclass[prl,amsfonts,english,showpacs,reprint,floatfix]{revtex4-1}
\usepackage[T1]{fontenc}
\usepackage[latin9]{inputenc}
\usepackage{graphicx}
\usepackage{esint}
\usepackage{verbatim}

\makeatletter
 
 \@ifundefined{textcolor}{}
 {%
   \definecolor{BLACK}{gray}{0}
   \definecolor{WHITE}{gray}{1}
   \definecolor{RED}{rgb}{1,0,0}
   \definecolor{GREEN}{rgb}{0,1,0}
   \definecolor{BLUE}{rgb}{0,0,1}
   \definecolor{CYAN}{cmyk}{1,0,0,0}
   \definecolor{MAGENTA}{cmyk}{0,1,0,0}
   \definecolor{YELLOW}{cmyk}{0,0,1,0}
 }

\makeatother

\begin{document}

\title{Mixed order phase transition in a one dimensional model}

\author{Amir Bar and David Mukamel}
\affiliation{Department of Complex Systems, Weizmann Institute, Rehovot, Israel}
\date{September 2, 2013}
\begin{abstract}
We introduce and analyze an exactly soluble one-dimensional Ising model with long range interactions which exhibits a mixed order transition (MOT), namely a phase transition in which the order parameter is discontinuous as in first order transitions while the correlation length diverges as in second order transitions. Such transitions  are known to appear in a diverse classes of  models which are seemingly unrelated. The model we present serves as a link between two classes of models which exhibit MOT in one dimension, namely, spin models with a coupling constant  which decays as the inverse distance squared and models of depinning transitions, thus making a step towards a unifying framework.

\end{abstract}
\pacs{64.60.De, 64.60.Bd, 05.70.Jk}

\maketitle
The usual classification of phase transitions distinguishes between
first order transitions which are characterized by a discontinuity
of the order parameter and second order transition in which the order
parameter is continuous but the correlation length and the susceptibility
diverge. However there are quite a number of cases for which this dichotomy
between first order and second order transitions fails. In particular,
some models exhibit phase transitions of mixed nature, which on the
one hand have a diverging characteristic length, as typical of second
order transitions, and on the other hand display a discontinuous order
parameter as in first order transitions. Examples include models of
wetting \cite{blossey1995diverging}, DNA denaturation \cite{PS1966,fisher1966effect,KMP2000},
glass and jamming transitions \cite{gross1985mean,schwarz2006onset,toninelli2006jamming,liu2012core},
rewiring networks \cite{liu2012extraordinary}
and some one dimensional models with long range interactions \cite{thouless1969long,dyson1971ising,aizenman1988discontinuity,luijten2001criticality}.
A scaling approach for such transitions was introduced in \cite{fisher1982scaling}. Formulating exactly
soluble models of this kind and probing their properties would be of great interest.

Two distinct classes of models which exhibit mixed transitions have
been extensively studied. (a) one dimensional spin models with interactions
which decay as $1/r^2$ at large distances $r$, and (b) models of DNA denaturation and
depinning transitions in $d=1$ dimension. While in both classes the
appropriate order parameter is discontinuous at the transition, the correlation length diverges
exponentially in the first class and algebraically in the second. Placing
the two rather distinct classes of models in a unified framework would provide a very interesting
insight into the mechanism which generates these unusual transitions. This is the aim of the present work.

An extensively studied representative of class (a) is the one dimensional Ising model with a ferromagnetic coupling which decays as  $1/r^\alpha$
with $\alpha=2$, which we shall call hereafter the inverse distance squared Ising (IDSI) model. While the model is not exactly soluble, many
of its thermodynamic features have been accounted for. It has been
shown by Dyson that for $1<\alpha<2$ the model exhibits
a phase transition to a magnetically ordered phase \cite{dyson1969existence}.
It has then been suggested by Thouless, and later proved rigorously by Aizenman et al. \cite{aizenman1988discontinuity}, that in the limiting case $\alpha=2$,
the model exhibits a phase transition in which the magnetization is
discontinuous \cite{thouless1969long}. This has been termed the
Thouless effect. Using scaling arguments \cite{anderson1969exact,anderson1970exact,anderson1971some} and renormalization group analysis \cite{cardy1981one}, which is closely related to the Kosterlitz-Thouless analysis, it was found that the correlation length diverges with an essential singularity as $\xi\sim e^{1/\sqrt{T-T_{c}}}$
for $T\rightarrow T_{c}$.

A paradigmatic example of models of class (b) is the Poland Scheraga (PS) model of DNA denaturation \cite{PS1966,fisher1966effect,KMP2000}
whereby the two strands of the molecule separate from each other at
a melting, or denaturation, temperature. In this approach the DNA molecule
is modeled as an alternating sequence of segments of bound pairs and
open loops. While bound segments are energetically favored, with an
energy gain $-\epsilon l$ for a segment of length $l$, an open loop
of length $l$ carries an entropy $sl-c\ln l$. Here $\epsilon,s>0$
are model dependent parameters and $c$ is a constant depending only on dimension and
other universal features. For $c>2$ the model has been shown to exhibit
a phase transition of mixed nature, with a discontinuity of the average
loop length which serves as an order parameter of the transition, and a
correlation length which diverges as $(T_{c}-T)^{-1}$ at the melting temperature $T_{c}$.

In this Letter we introduce and study an exactly soluble variant of the IDSI
model in which the the $1/r^2$ interaction applies only to spins which lie in the same domain of either up or down spins.
This model can be conveniently represented within the framework of the Poland Scheraga model, thus providing a link between these broadly studied classes of models.
We find that on one hand the model exhibits an extreme Thouless effect
whereby the magnetization $m$ jumps from $0$ to $\pm 1$ at $T_{c}$,
and on the other hand it exhibits an algebraically diverging correlation
length $\xi\sim\left(T-T_{c}\right)^{-\nu}$, and consequently a diverging
susceptibility. The power $\nu$ is model dependent and it varies with the model parameters.
We also identify an additional order parameter, the average number of domains per unit length, $n$, which
vanishes either continuously or discontinuously at the transition,
depending on the interaction parameters of the model. In addition we find a similar type of transition (discontinuous with diverging correlation length) at non-zero magnetic field. This is in contrast to the IDSI model which exhibit no transition for non-vanishing magnetic field \cite{fisher1982scaling}.
Below we demonstrate these results by an exact calculation. We also present an RG analysis which provides a common framework for studying both the IDSI and our model, elucidating the relation between the two.

The model is defined on a one dimensional lattice with $L$ sites
where in site $i$, $1\le i\le L$, the spin variable $\sigma_{i}$
can be either $1$ or $-1$. The Hamiltonian of the model is composed
of two terms: nearest neighbor (NN) ferromagnetic term $-J_{NN}\sigma_{i}\sigma_{i+1}$
and a long range (LR) term which couples spins lying within the same
domain of either up or down consecutive spins. This intra-domain interaction
is of the form $-J(i-j)\sigma_{i}\sigma_{j}$ where $J\left(r\right)$
decays as $J_{LR}r^{-2}$ for large $r$. This is a truncated version
of the IDSI model. Note, though, that the LR interaction is in fact
a multi-spin interaction since it couples only spins which lie in
the same domain. For domains of length $l\gg1$ the energy due to the intra-domain interactions
is
\begin{eqnarray}
E_{d}\left(l\right) & \approx & -J_{NN}(l-1)-J_{LR}\sum_{k=1}^{l}\frac{l-k}{k^{2}}\nonumber \\
 & = & -bl+\tilde{c}\log l+\tilde{\Delta}+O\left(l^{-1}\right),\label{eq:Ed_def}
\end{eqnarray}
where $b,\tilde{c}$ and $\tilde{\Delta}$ are constants set by $J_{NN}$
and $J_{LR}$. Without loss of generality one may set $b=0$ since it contributes
a constant to the total energy.
Nearest neighbor domains interact only through the $NN$ interaction. The interaction
$J_{NN}>0$ can be made large enough so that $\Delta\equiv J_{NN}+\tilde{\Delta}>0$
and hence domain walls are disfavored and the model is ferromagnetic.
A configuration of the model is composed of a sequence of $N$ domains of alternating signs
whose lengths $\left\{ l_{i}\right\} _{i=1}^{N}$ satisfy
$\sum l_{i}=L$. The corresponding energy is
\begin{equation}
H\left(\left\{ l_{i}\right\},N \right)=\tilde{c}\sum_{i=1}^{N}\log\left(l_{i}\right)+N\Delta+O(1).\label{eq:hamiltonian}
\end{equation}
  This representation of the model is reminiscent of the PS model, where
$E_{d}(l)$ originates from the \emph{entropy }of a denatured loop rather than its energy \cite{PS1966}. We also generalize (\ref{eq:hamiltonian}) to include a magnetic field $h$, which couples to the magnetization $\sum_i(-1)^i l_i$.

\begin{figure}
\includegraphics[scale=0.6]{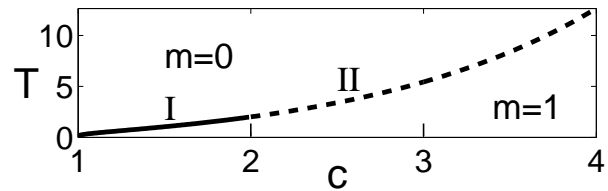}
\caption{\label{fig:Phase_diagram}The ($T$,$c$) phase diagram of the model (\ref{eq:hamiltonian}) for zero magnetic field $h=0$ and $\Delta=1$.
 The transition line is marked as a continuous and dashed line in regions I $(1<c\le 2)$ and II $(c>2)$, respectively}
\end{figure}

This model is exactly solvable. The phase diagram of the model at zero magnetic field is presented in Fig.\ref{fig:Phase_diagram}. The model exhibits a phase transition from a disordered phase ($m\equiv\sum_i^L \sigma_i/L=0$) at $T>T_c$ to a fully ordered phase ($m=\pm 1$) at $T<T_c$, where $T_c$ is the critical temperature. While $m$ is discontinuous at the transition, the correlation length diverges and hence the transition is of mixed order. In addition to the magnetization, the transition may be characterized by another order parameter, the density of domains, $n\equiv N/L$. In the disordered phase there is a macroscopic number of domains and hence $n>0$, while in the ordered phase there is essentially a single macroscopic domain --- \emph{a condensate} --- and hence $n=0$. The behavior of $n$ near the transition depends on the non-universal parameter $c\equiv\beta_{c}\tilde{c}$ where $\beta_{c}=(k_{B}T_{c})^{-1}$ is the inverse transition temperature which depends on the interaction parameters: For $1<c\le2$ (region $I$ in Fig.\ref{fig:Phase_diagram}) the density of domains decreases continuously to $0$ as $T\rightarrow T_{c}$ from above, while for $c>2$ (region $II$), $n$ attains a finite value $n\rightarrow n_c$ as $T\rightarrow T_c$ from above, and it drops discontinuously to $0$ at the transition.

\begin{figure}
\includegraphics[scale=0.6]{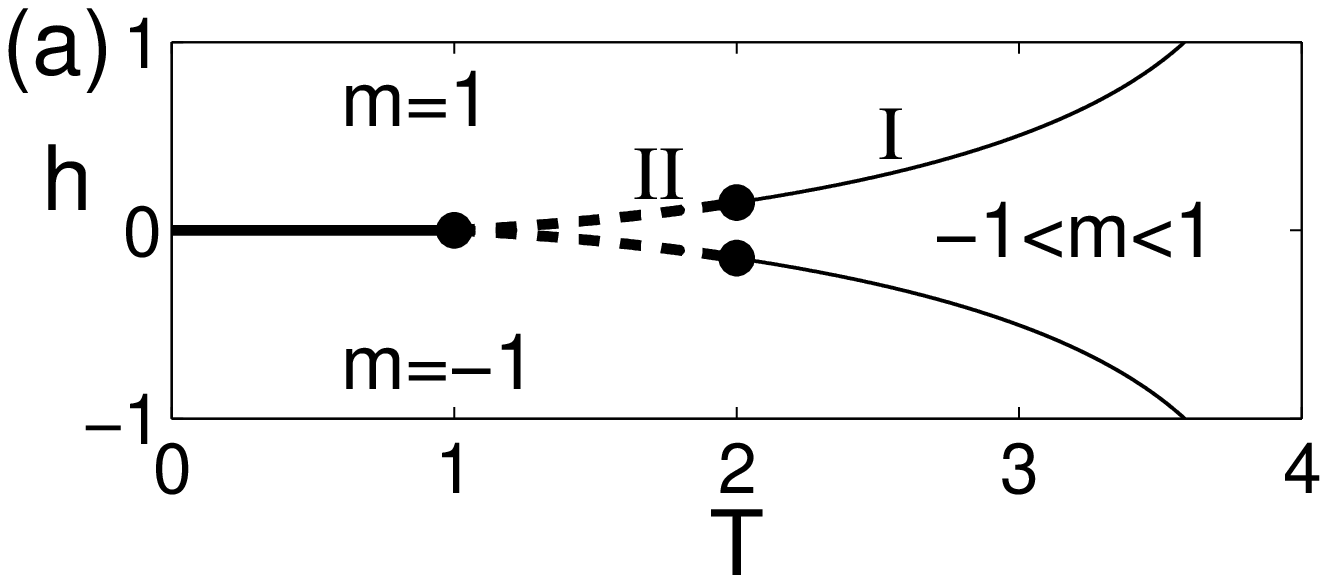} \\
\includegraphics[scale=0.6]{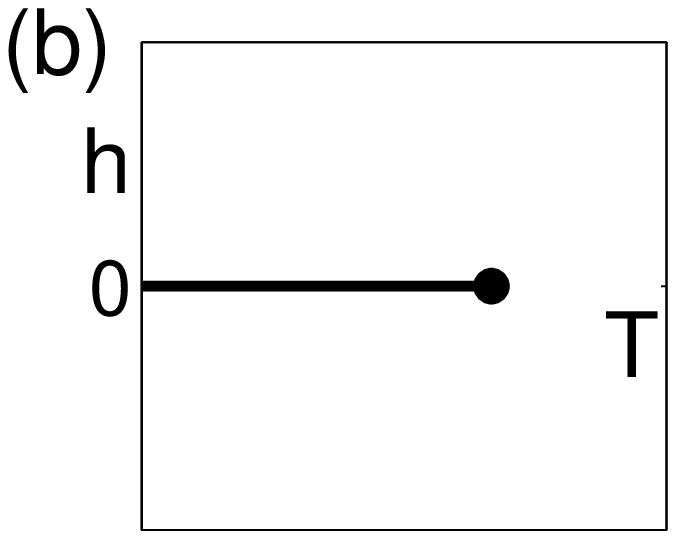} \includegraphics[scale=0.6]{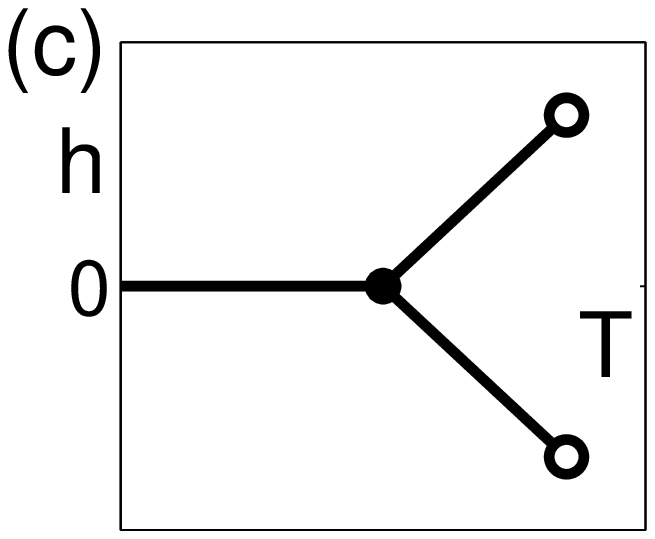}

\caption{\label{fig:mag_pt} (a) The ($T$,$h$) phase diagram of the model (\ref{eq:hamiltonian}) compared with the phase diagram of (b) the IDSI model and (c) a schematic phase diagram of a typical first order transition.
The parameters in (a) are $\tilde{c}=4$ and $\Delta=\log(U(0))$ so that $T_c(h=0)=1$. Here the thick solid line is first order transition, the dashed lines represent mixed order transition, and the thin solid lines a second order transition. Tricritical points separating mixed order from second order transitions are indicated. In (b) and (c) the lines
are first order. The terminal point in (b) is a mixed order point. In (c) the solid point is a triple point while
the empty circles are ordinary critical transitions.}
\end{figure}

The model also exhibits a condensation transition at finite magnetic field as presented in Fig.\ref{fig:mag_pt}a. The transition at non-zero field does not involve symmetry breaking. It can be either second order, where both $m$ and $n$ change continuously to their ordered values $m=\pm1$ and $n=0$, or of mixed order, where both $m$ and $n$ change discontinuously at the transition. This depends on whether $c(h)\equiv \beta_c(h)\tilde{c}$ is greater or smaller than $2$, where $\beta_c(h)$ is the magnetic-field dependent critical temperature. Qualitatively the phase diagram at a given magnetic field $h\neq0$ is identical to that of the PS model, with $c(h)$ playing the role of $c$. On the other hand, the resulting phase diagram (Fig.\ref{fig:mag_pt}a) is different from that of the IDSI model, presented in Fig.\ref{fig:mag_pt}b, for which no transition takes place at a non-vanishing magnetic field \cite{fisher1982scaling}. It is also different from the phase diagram of ordinary first order transition such as the mean-field Ising spin 1 model, which is presented in Fig.\ref{fig:mag_pt}c, for which each of the finite $h$ transition lines terminates at a critical point at some finite value of $h$. By contrast, the finite $h$ transition lines in Fig.\ref{fig:mag_pt}a extend to $h\rightarrow\infty$.

We shall now outline the derivation of the phase diagram. The Hamiltonian (\ref{eq:hamiltonian}) represents a gas of non-interacting domains with a fugacity $\Delta$. Correlation between domains is introduced, though, by the constraint that the sum of $l_i$ is $L$, the chain length. The system is thus most conveniently studied within the grand canonical (GC) ensemble. The GC partition function is given by
\begin{eqnarray}
Q(p)&=&\sum_{L}Z\left(L\right)e^{\beta pL} \simeq \sum_{L}e^{-\beta L\left(F(\beta) - p\right)}.\label{eq:Q_def}
\end{eqnarray}
where $Z\left(L\right)$ is the canonical partition function, $F(\beta)$ is the free energy per site and $p$ is effectively the pressure. Working with the symmetric boundary conditions $\sigma_{1}=1$ and $\sigma_{L}=-1$, a configuration is defined by a sequence of an even number of alternating $+$ and $-$ domains of variable sizes. Denoting by $U(p)$ the grand partition sum of a single domain and by $y\equiv e^{-\beta\Delta}$ the fugacity of domains, the explicit form of the grand partition sum is then

\begin{eqnarray}
Q(p) & \sim & y^2 U(p)^2 + y^4 U(p)^4+... = \frac{y^2U(p)^2}{1-y^2U(p)^2},\label{eq:Qzqr}\\
U(p) & = & \sum_{l=1}^{\infty}\frac{e^{\beta p l}}{l^{\beta\tilde{c}}}=\Phi_{\beta\tilde{c}}\left(e^{\beta p}\right),\label{eq:U_def}
\end{eqnarray}
where  $\Phi_{\gamma}(r)$ is the polylogarithm function \cite{Lewin1981}. Using the properties of the polylogarithm, or just inspecting the sum in (\ref{eq:U_def}), we see that for $p\le0$,  $U(p)$ is an increasing function of $p$, a decreasing function of $\beta$ and has a branch point at $p=0$. In the thermodynamic limit $L\rightarrow\infty$, the most negative singularity of $Q(p)$ is given by $p^*=F(\beta)$ as this sets the radius of convergence of the sum in (\ref{eq:Q_def}). The singularity can stem either from setting the denominator of (\ref{eq:Qzqr}) to zero, or from the branch-point of $U(p)$, i.e.
\begin{equation}
(a)\; U\left(p^*\right)=e^{\beta\Delta}\quad or \quad (b)\; p^*=0 \label{eq:ps_cond}
\end{equation}
The solution of (a) corresponds to the state with zero magnetization (no condensate) while (b) corresponds to the magnetic state.
At high enough temperatures for which $\beta \tilde{c}<1$ the sum $U(0)=\sum l^{-\beta\tilde{c}}$ diverges and a solution of type (a), with $p^*<0$, exists. The solution $p^*$ increases with increasing $\beta$ and at the critical point $\beta_c$, for which $c=\beta_c \tilde{c}>1$ and hence $U(0)<\infty$, $p^*$ vanishes. It stays zero at all temperatures below $T_c$. Thus $\beta_c$ is a singular point of the free energy $F(\beta)=p^*$. The freezing of the thermodynamic pressure $p^*$ below $T_c$ is mathematically similar to the freezing of the fugacity in Bose-Einstein Condensation (BEC) for free bosons \cite{Hu1987}.

We next proceed to show that there is a diverging length scale. Above the transition the probability to have a domain of size $l$ is given by
\begin{equation}
P(l)\simeq\frac{Z(L-l)yl^{-\beta\tilde{c}}}{Z(L)}\simeq y\frac{e^{-l/\xi}}{l^{\beta\tilde{c}}},\label{eq:Pl}
\end{equation}
where we have used the fact that $Z(L)\simeq e^{-\beta p^* L}$, and defined $\xi=-\left(\beta p^*\right)^{-1}$. The
length scale $\xi$ can be regarded as a correlation length, and it
diverges at the transition (for any $c$) as $p^*\rightarrow 0$. Expanding Eq.(\ref{eq:ps_cond}a) near the transition, it can be shown that $\left(-p^{*}\right)^{\min\left(c-1,1\right)}\sim\left(T-T_{c}\right)$.
Hence we deduce that $\xi\sim\left(T-T_{c}\right)^{-\nu}$ with $\nu=\max\left(\frac{1}{c-1},1\right)$, demonstrating the algebraic divergence of the correlation length for all $c>1$.

The average density of domains is given by the usual relation $\left<n\right>=-\frac{\partial{p^*}}{\partial \Delta}$. From this it is easy to see that at the low temperature phase $\left<n\right>=0$ since $p^*=0$ regardless of $\Delta$. As $\left<n\right>\times\left<l\right>=1$ this implies that $\left<l\right>=\infty$ for $T<T_c$. At the transition, where the correlation length $\xi$ diverges, the average domain length is given by $\left<l\right>=\sum_l lP(l)=\sum l^{-c+1}$ and hence it is finite if $c>2$ and infinite if $1<c\le2$. This implies that $\left<n\right>$ drops continuously to $0$ if $1<c\le2$ and discontinuously if $c>2$.

Finally we wish to show that the magnetization jumps at $T_c$ from $0$ to $\pm1$ for all $c>1$. At zero magnetic field the system has spin reversal symmetry and hence as long as the symmetry is not spontaneously broken (i.e. at the high temperature phase) the magnetization is $0$. The low temperature phase is characterized by a condensate, as was argued by the similarity to BEC and also as $\left<n\right>=0$, i.e. there is essentially a single macroscopic domain (plus maybe a sub-extensive number of microscopic domains). As the condensate is either of type $+1$ or $-1$, we find $\left<m\right>=\pm1$. This demonstrates the features of the phase diagram shown in Fig.\ref{fig:Phase_diagram}.

We now consider the finite magnetic field case. The analysis of the transition in this case follows essentially the same steps as for the zero magnetic field case, with Eq.(\ref{eq:ps_cond}) replaced by
\begin{equation}
(a)\; U\left(p^*+h\right)U\left(p^*-h\right)=e^{2\beta\Delta} \;\; or \;\; (b)\; p^*=-|h|. \label{eq:p_cond}
\end{equation}
At finite $h$, the magnetization $m=-\frac{\partial p^*}{\partial h}$ is non-zero even in the high temperature phase. For $1<c\le 2$ it is continuous at $T_c$ and the transition is an ordinary second order transition. For $c>2$, $\left<m\right>$ is discontinuous at $T_c$ and the transition is of mixed nature as depicted in Fig.\ref{fig:mag_pt}a.

\begin{figure}
\includegraphics[scale=0.3]{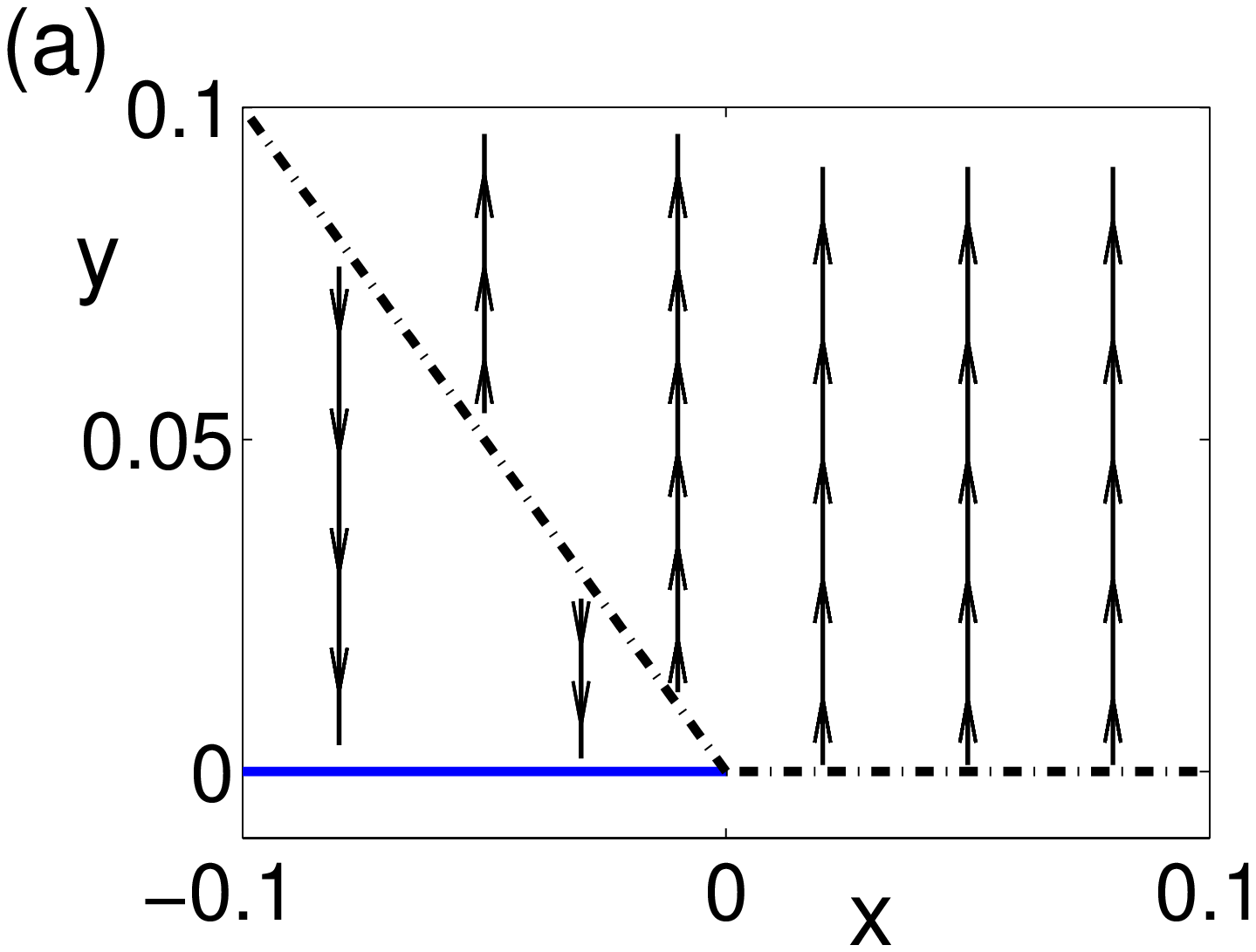}\includegraphics[scale=0.3]{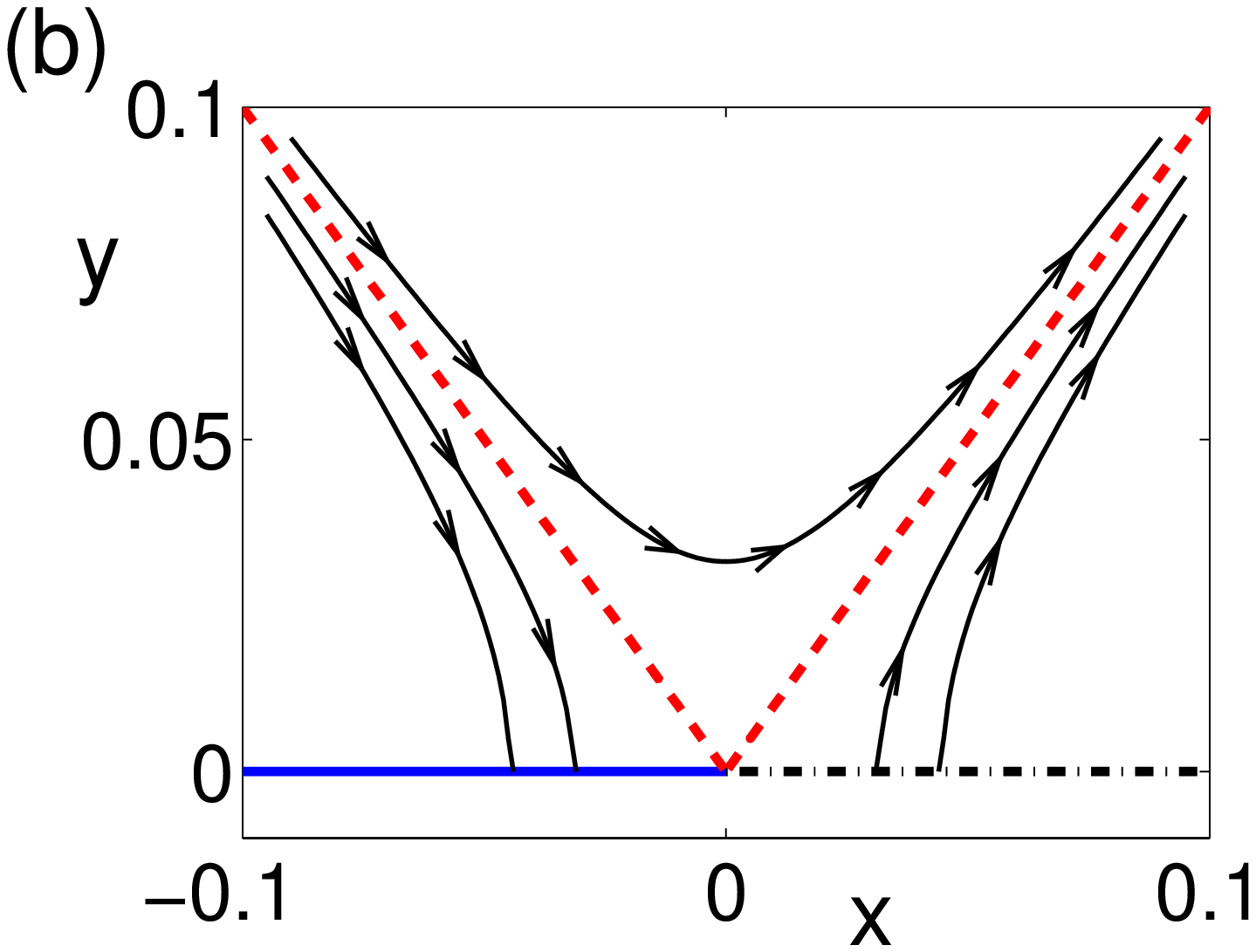}

\caption{\label{fig:RG-Flow} (color online) RG flow for (a) the truncated model, Eq.(\ref{eq:rg_flow})
and (b) the IDSI model (or XY model), Eq.(\ref{eq:KT_RG}). Solid lines indicate attractive fixed points, unstable fixed points are marked by dashed-dotted lines, and the dashed line in (b) is a separatrix. }
\end{figure}

It is instructive to consider the renormalization group (RG) flow
of the model and compare it with that of the IDSI model. This provides a common analytical framework for both models and help elucidating the mechanism behind their distinct features.
The RG flow of the IDSI model has been studied first by Anderson
et al.\cite{anderson1969exact,anderson1970exact,anderson1971some}
using scaling arguments and then more systematically by Cardy \cite{cardy1981one}
and was shown to be of the Kosterlitz-Thouless type \cite{kosterlitz1973ordering}.
In particular the transition is characterized by a length which diverges
as $\exp\left[\left(T-T_{c}\right)^{-1/2}\right]$. We show below that
in our model the RG equations are of different form, yielding a correlation
length which diverges with a power law. To proceed
we consider a continuous version of the model, which captures the long wavelength behavior of the original model: We represent the domain
boundaries (the kinks) as particles with impenetrable core of size
$a$, placed on a circle at positions $\left\{ r_{i}\right\} _{i=1}^{N}$,
whereby, following Eq.(\ref{eq:Ed_def}), every pair of nearest neighbor particles $i$ and $i+1$ attract
each other logarithmically through a two body potential $\tilde{c}\log\left(r_{i+1}-r_{i}\right)$. The number of particles is not
conserved, as the number of kinks in the spin representation fluctuates, and
it is controlled by a fugacity $y$ (equivalent to $e^{-\beta\Delta}$ above). The
partition function is thus
\begin{equation}
Z=\sum_{N=0}^{\infty}y^{N}\int\prod_{i=1}^{N}\frac{dr_{i}}{a}\left(\frac{r_{i+1}-r_{i}}{a}\right)^{-\beta \tilde{c}}\Theta\left(r_{i+1}-r_{i}-a\right),\label{eq:cardy_Z}
\end{equation}
where $\Theta$ is the Heaviside step function. Assuming small density
of particles ($y\ll1$), the renormalization procedure proceeds by
rescaling the core size of the particle $a\rightarrow ae^{\kappa}$, as
in \cite{cardy1981one}. The resulting flow equations in terms of
the fugacity $y$ and the scaled interaction strength $x=1-\beta\tilde{c}$
read
\begin{equation}
\frac{dy}{d\kappa}=xy+y^{2}\quad;\quad\frac{dx}{d\kappa}=0.\label{eq:rg_flow}
\end{equation}
The $xy$ term in (\ref{eq:rg_flow}) compensates for the
change in the $a^{-\left(N-N\beta\tilde{c}\right)}$ factor in (\ref{eq:cardy_Z}),
and is the same as in the analysis of the IDSI model \cite{cardy1981one}.
The second term ($y^{2}$) is the result of expanding the $\Theta$
function as $\Theta\left(r_{i+1}-r_{i}-ae^{\kappa}\right)\approx\Theta\left(r_{i+1}-r_{i}-a\right)-a\kappa\delta\left(r_{i+1}-r_{i}-a\right)$.
Physically the second term of this expansion corresponds to the merging
of two kinks due to the rescaling procedure, and hence it results in
the $y^{2}$ term. As these are the
only effects of the scale transformation, $x$ remains invariant
under it. The resulting flow diagram is presented in Fig. \ref{fig:RG-Flow}a.
In this flow there is a line of unstable fixed points for $y=-x$ each corresponding to a
different value of $c$. Similar flow diagram has previously been found for the one dimensional
discrete gaussian model with $1/r^2$ coupling \cite{slurink1983roughening,guinea1985diffusion}.

Equations (\ref{eq:rg_flow}) can be compared with the RG equations
for the IDSI model which are the same as those of the XY model (under proper
rescaling of parameters) \cite{cardy1981one,kosterlitz1973ordering}
\begin{equation}
\frac{dy}{d\kappa}=xy\quad;\quad\frac{dx}{d\kappa}=y^{2}.\label{eq:KT_RG}
\end{equation}
Notice that in this case the merging of two
kinks produces a dipole interaction,
and hence the $y^2$ term renormalizes the interaction strength $x$. The renormalization flow of this model is presented in Fig.\ref{fig:RG-Flow}b.
It has only a single unstable fixed point for $x=y=0$ in the
relevant $x\le0$ regime.

One can calculate the temperature dependence of the correlation length of the truncated model
by linearizing Eq.(\ref{eq:rg_flow}) near the fixed points.
The result is $\xi\sim\left[\left(T-T_{c}\right)/\left|x\right|\right]^{\frac{1}{x}}$,
which is the same as that found above for $c\le2$.

In conclusion, we have presented and analyzed a novel one dimensional Ising
model which displays a spontaneous symmetry breaking transition with
diverging correlation length and an extreme Thouless effect, i.e. a discontinuous jump in magnetization
(from $0$ to $\pm1$).
The model conveniently connects two widely studied classes of models, the Poland Scheraga model and the IDSI.
In addition to the magnetization we have identified another order parameter, the density of domains, $n$,
and showed that it is either continuous or discontinuous at $T_{c}$
depending on whether $c\le2$ or $c>2$, respectively. This order
parameter has not been discussed in the context of the IDSI model, and it would be interesting to explore its behavior in that case. We also showed that the model exhibits mixed transitions for non-zero magnetic field, unlike the IDSI model, and hence it does not fall into the classification of first order transition points appearing in \cite{fisher1982scaling}.
We have also used an RG picture to explain the power law divergence
of the correlation length in this model, in contrast to the essential
singularity behavior of the correlation length in the IDSI model. It would be interesting
to extend the present study to Potts type models and to consider the effect of disorder on the nature of the transition.


We thank M. Aizenman, O. Cohen, O. Hirschberg and Y. Shokef for helpful discussions. The support of the Israel Science Foundation
(ISF) and of the Minerva Foundation with funding from the Federal German Ministry
for Education and Research is gratefully acknowledged.
\bibliographystyle{h-physrev}

\end{document}